\documentclass[aps,prl,twocolumn,groupedaddress,showpacs]{revtex4}

\usepackage{graphicx}

\newcommand{\an}{\hat{a}}
\newcommand{\bn}{\hat{b}}
\newcommand{\cn}{\hat{c}}

\newcommand{\bd}{\hat{b}^\dagger}

\newcommand{\Geff}{G_\mathrm{eff}}

\begin{document}

\title{Strong Low-Frequency Quantum Correlations From a Four-Wave Mixing Amplifier}

\author{C. F. McCormick}
\author{A. M. Marino}
\author{V. Boyer}
\author{P. D. Lett}\email{paul.lett@nist.gov}
\affiliation{Atomic Physics Division, National Institute of
Standards and Technology, Gaithersburg MD 20899 USA }

\date{\today}

\begin{abstract}
We show that a simple scheme based on nondegenerate four-wave mixing
in a hot atomic vapor behaves like a near-perfect phase-insensitive
optical amplifier, which can generate bright twin beams with a
measured quantum noise reduction in the intensity difference of more
than 8 dB, close to the best optical parametric amplifiers and
oscillators.  The absence of a cavity makes the system immune to
external perturbations, and the strong quantum noise reduction is
observed over a large frequency range.
\end{abstract}


\maketitle

Two-mode squeezed beams have become a valuable source of entanglement
for quantum communications and quantum information
processing~\cite{Braunstein2005}.  These applications bring specifi
requirements on the squeezed light sources.  For instance, for squeeze
light to be used as a quantum information carrier interacting with 
material system, as in an atomic quantum memory, the
light field must be resonant with an atomic transition and spectrally
narrow to ensure an efficient coupling between light and matter. In
recent years, attention has also been brought to the problem of the
manipulation of cold atomic samples with non-classical fields in order
to produce non-classical matter
waves~\cite{Lett2004,Haine2005,Haine2006}. In this case, the slow
atomic dynamics also requires squeezing at low frequencies.

The standard technique for generating nonclassical light fields is by
parametric down-conversion in a crystal, with an optical parametric
oscillator or an optical parametric
amplifier~\cite{Heidmann1987,Feng04}. While very large amounts of
quantum noise reduction have been achieved in this
way~\cite{Laurat2005,Vahlbruch2007}, controlling the frequency and the
linewidth of the light remains a challenge. Only recently have sources
based on periodically-poled nonlinear crystals been developed at
795 nm to couple to the Rb D1 atomic
line~\cite{Tanimura2006,Hetet2006}. On the other hand, stimulated
four-wave mixing (4WM) naturally generates narrow-band light close to
an atomic resonance, but its development as an efficient source of
squeezed light has been hindered by fundamental limitations such as
spontaneous emission. At the end of the 1990s, nondegenerate 4WM in a
double-lambda scheme was identified as a possible workaround for these
limitations, as described in Ref.~\cite{Lukin2000} and references
therein. It was not until recently that such a scheme was implemented
in continuous mode in an efficient way in both the
low~\cite{Balic05,Kolchin2006,Thompson2006} and the
high~\cite{vanderWal2003,McCormick2007} intensity regimes, where it was
shown to generate twin beams where quantum correlations are not masked
by competing effects.

The double-lambda scheme gives rise to complex atomic dynamics and
propagation properties, such as slow-light effects~\cite{Boyer2007}.
In this Letter, we show that in spite of this complexity, the
quantum properties of the scheme can be accurately described as the
combination of a perfect amplifier and a partial absorber.  This
model allows us to optimize the quantum noise reduction in the
intensity difference of the bright twin beams and to isolate the
limiting factors of this reduction. It also helps to identify
regions of the parameter space where the system behaves like a
perfect phase-insensitive amplifier, opening the way to the
generation of strong continuous-variable entanglement. Finally, we
demonstrate the intrinsic robustness of our scheme by measuring
large levels of squeezing in the audio range, at frequencies where
technical noise usually represents a serious obstacle to the
generation and the observation of quantum effects.

The double-lambda scheme (Fig. 1a) is a 4WM process which,
\emph{via} the interaction with 4 atomic levels, mixes 2 strong pump
fields with a weak probe field in order to generate a fourth field
called the conjugate.  The probe and conjugate fields (the twin
beams) are cross-coupled and are jointly amplified, which leads to
intensity correlations stronger than the standard quantum limit
(SQL).  These correlations are the manifestation of two-mode
quadrature squeezing between opposite vacuum sidebands of the twin
beams.

\begin{figure}[ht]
\centering
\includegraphics[width=\columnwidth]{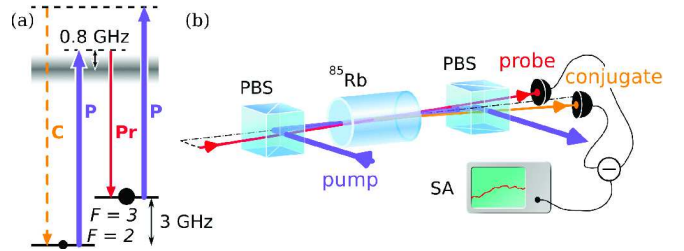}
\caption{(color online).  Experimental details. (a) Four-level
double-lambda scheme in $^{85}$Rb, P = pump, C = conjugate, Pr =
probe.  The width of the excited state represents the Doppler
broadened profile. (b) Experimental setup, PBS = polarizing beam
splitter, SA = spectrum analyzer. \label{Setup}}
\end{figure}

As in the experiments presented in \cite{McCormick2007}, we use a cw
Ti:Sapphire ring laser, to generate a strong ($\approx$~400~mW) pump
beam near the D1 line of Rb (795 nm).  From this we derive, using an
acousto-optic modulator, a weak ($\approx$~100~$\mu$W) probe beam
tuned $\approx$ 3~GHz to the red of the pump. This results in very
good relative phase stability of the probe with respect to the pump.
The pump and probe beams are cross-linearly polarized, combined in a
Glan-Taylor polarizer, and directed (at an angle of 0.3 degrees to
each other) into a 12.5~mm vapor cell filled with isotopically pure
$^{85}$Rb (see Fig.~\ref{Setup}). The cell, with no magnetic
shielding, is heated to $\approx 110^\circ$C. The windows of the
cell are anti-reflection coated on both faces, resulting in a
transmission for the probe beam of 98\% per window. The pump and
probe are collimated with waists at the cell position of 650 $\mu$m
and 350 $\mu$m (1/e$^2$ radius) respectively.

After the cell we separate the pump and probe beams using a second
polarizer, with $\approx 10^5:1$ extinction ratio for the pump. With
the pump at $\omega_0$, tuned to a `one-photon detuning' of 800~MHz
to the blue of the $^{85}$Rb $5S_{1/2} F=2 \rightarrow 5P_{1/2}$, D1
transition, and the probe at $\omega_-$, detuned 3040 MHz to the red
of the pump (`two-photon detuning' of 4~MHz), we measure an
intensity gain on the probe of 9. This gain is accompanied by the
generation of the conjugate beam at $\omega_+$, detuned 3040 MHz to
the blue of the pump, which has the same polarization as the probe,
and propagates at the pump-probe angle on the other side of the pump
so that it fulfills the phase-matching condition. After the second
polarizer we direct the probe and conjugate beams into the two ports
of a balanced, amplified photodetector. The output of this
photodetector is fed into a radio frequency spectrum analyzer with a
resolution bandwidth (RBW) of 30 kHz and a video bandwidth (VBW) of
300 Hz. In addition, we introduce a delay line into the conjugate
beam path to compensate for the differential slow-light delay
discussed in Ref.  \cite{Boyer2007}.  This results in a fraction of a
dB improvement in the amount of squeezing observed and increases the
squeezing bandwidth up to 20~MHz.

\begin{figure}[ht]
\includegraphics[width=0.8\columnwidth]{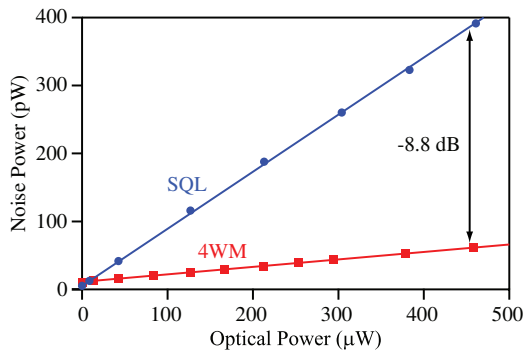}
\caption{(color online). Intensity-difference noise versus total
optical power at 1~MHz. Circles: SQL; red squares: 4WM. The
ratio of the two slopes is $-8.8$~dB.
\label{NoiseVPower}}
\end{figure}

We measure $-8.0$~dB of intensity-difference squeezing at an analysis
frequency of 1~MHz without compensating for any system noise.  This
noise has contributions from the electronic noise of the detection
system and the pump light that scatters from the atomic medium. In
order to determine the effect of the system noise on the measured
squeezing we vary the input probe power and we plot in
Fig.~\ref{NoiseVPower} the intensity-difference noise of the probe
and conjugate beams versus the total (probe plus conjugate) power
incident on the detector, as well as the standard quantum limit
(SQL), determined by measuring shot-noise-limited balanced beams of
the same total power. The two curves fit to straight lines, with a
ratio of slopes equal to 0.131 = $-8.8$~dB. The SQL curve has a
zero-intercept given by $-82.9$~dBm, while the zero-intercept of the
probe-conjugate curve is higher, $-79.6$~dBm (due to the pump
scattering). The optical path transmission and photodiode
efficiencies are (95.5 $\pm$ 2)\% and (94.5 $\pm$ 2)\%,
respectively, resulting in a total detection efficiency of $\eta$ =
(90 $\pm$ 3)\%; all uncertainties are estimated 1 standard
deviation. The squeezing value at the end of the atomic medium,
corrected for losses, is better than $-11$~dB.

We turn to a simple model of distributed gain and loss in the
medium~\cite{Jeffers1993,Kumar2002} to isolate the few physical
concepts necessary to describe the 4WM process and to quantitatively
explain the measured squeezing.  Two-mode squeezing is produced in an
ideal medium of gain $G$, with negligible absorption, where the photon
annihilation operators $\an$ and $\bn$ for the probe and the conjugate
fields transform according to
\begin{eqnarray}
    \an^{\phantom{\dagger}} &\rightarrow& \an\sqrt{G} - \bd\sqrt{G - 1}
    \label{gain1}\\
    \bd &\rightarrow& \bd\sqrt{G}-\an\sqrt{G - 1}   \label{gain2}.
\end{eqnarray}
  
When the probe port $a$ is seeded with a
coherent state $|\alpha\rangle$ and the conjugate port $b$ is fed
with the vacuum, the output intensity-difference noise is equal to
the shot noise of $|\alpha\rangle$, which gives a quantum noise
reduction of $1/(2G - 1)$ with respect to the output SQL.

In our experiment the probe, unlike the conjugate, is tuned close
enough to an atomic resonance to experience some absorption [see Fig.
\ref{Setup}(a)]. The coherent coupling between probe and pump leads to
a certain degree of electromagnetically-induced transparency (EIT) for
the probe \cite{Lukin2000}.  Imperfections such as stray magnetic
fields, the residual Doppler effect, atomic collisions and short
atomic transit time through the beams, as well as the depumping due t
the conjugate-pump lambda system limit this effect. In
addition, the atomic susceptibility results in an offset of the gain
maximum from the EIT absorption minimum \cite{Boyer2007}. Localized
loss in the system can be modeled by a beam splitter of transmission
$T$ picking off a fraction of the probe and injecting the vacuum on
the second input port (photon annihilation operator $\cn$), according
to the transformation:
\begin{eqnarray}
    \an^{\phantom{\dagger}} &\rightarrow& \an\sqrt{T} +
    \cn\sqrt{1 -T} \label{split1}\\
    \bd &\rightarrow& \bd \label{split2}.
\end{eqnarray}
Because of the vacuum noise injected into port $c$, such a
transformation applied to the squeezed probe and conjugate is expected
to degrade the squeezing. In the actual medium, gain and loss are
distributed and can be modeled by a succession of $N\gg1$ interleaved
stages of elementary gain $g$ and $N$ stages of elementary transmission
$t$. The intrinsic gain $G$ (probe transmission $T$) of the whole stack is
obtained by making $t=1$ ($g=1$), respectively. Finally, the total
detection efficiency $\eta$ is also modeled by a beam splitter of
transmission $\eta$ applied to both the probe and conjugate fields. In
the calculations below we set $N = 200$.

\begin{figure}[ht] 
    \includegraphics[width=\columnwidth]{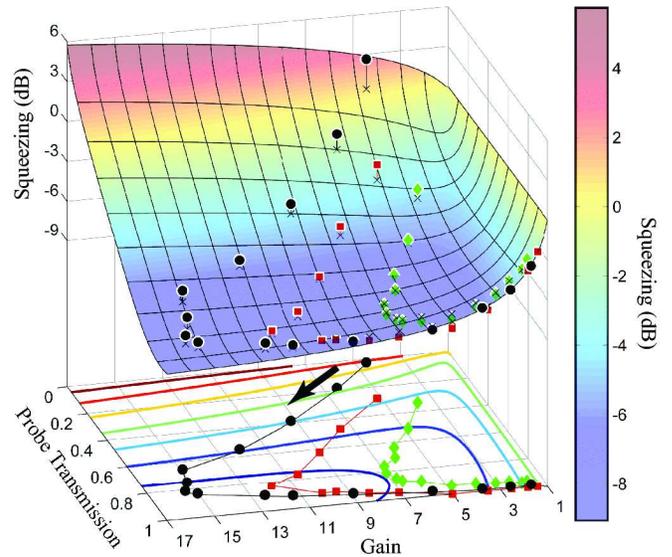}
    \caption{\label{model}(color online). Simulated and measured
    intensity-difference squeezing as a function of the probe
    transmission $T$ and medium gain $G$. The theory takes into
    account the detection efficiency ($\eta$ = 0.9). The squeezing
    (corrected for the system noise) measured at 1 MHz is shown for
    different cell temperatures, 109$^{\circ}$C (diamonds),
    112$^{\circ}$C (squares), and 114$^{\circ}$C (circles), as the
    one-photon detuning of the pump laser is scanned. The crosses
    indicate the projection of the measured squeezing onto the
    theoretical surface while the lines connecting the spheres and
    crosses give an indication of the vertical distance between them.
    Most of the points are very near the surface. The projection
    onto the $x-y$ plane shows contour lines of the theoretical
    squeezing at 2 dB intervals from +4 to $-8$ dB, and the
    projections of the data points. The arrow indicates the direction
    of increasing one-photon detuning.} 
\end{figure}

The experimentally accessible parameters are the effective (measured)
probe gain of the medium $\Geff$, defined as the probe power out of
the cell divided by the probe power in (corrected for window losses),
and the ratio $r$ of conjugate-to-probe output powers. Both parameters
depend on the intrinsic gain the loss of the probe in the medium.
From $\Geff$ and $r$, we use the model to numerically determine $G$
and $T$ without any free parameters.  Experimentally, the parameter
space is probed by scanning the one-photon detuning from 0.4 to
1.4~GHz.  This effectively changes both the gain and the transmission
of the probe in a coupled manner.  Figure~\ref{model} shows the
theoretical intensity-difference squeezing obtained from the model
described above as well as the measured squeezing (corrected for the
system noise) as a function of the probe transmission $T$, and the
gain $G$. The main result is that, as shown in Fig.~\ref{model}, the
experimental data points agree very well with the simple gain/loss
model.

The theoretical surface of Fig.~\ref{model} shows that for a given
probe transmission there is an optimum gain.  At lower gain the
intensity-difference squeezing is limited by the power imbalance
between the probe and the conjugate, which originates from the input
probe power.  At larger gain the squeezing becomes limited by the
amplification of the noise introduced by the loss on the probe.  In th
same way,
for a given gain there is an optimum transmission value, smaller than
1, corresponding to this trade-off between power balancing and
absorption-induced quantum noise.  In order to increase the squeezing,
it would be necessary to both reduce the probe absorption and increase
the gain.  Experimentally, the absorption and the gain both depend on
the cell temperature (both increase with the atomic density) and the
pump detuning, and they are not independently controllable.  The best
squeezing of $-8.8$~dB is obtained over the transmission range of
0.85--0.95, and gains of 9--15. This best value can be achieved over a
range of several degrees in temperature.

An interesting feature revealed by the data in Fig.~\ref{model} is
that at large one-photon detunings, the probe transmission becomes
unity for an intensity-difference squeezing of about $-7$~dB.  It is
therefore possible to operate the system as an ideal amplifier with a
gain up to about 6, producing, in principle, a pure entangled state
which could be a valuable resource for some quantum information
protocols.

\begin{figure}[ht] 
    \includegraphics[width=\columnwidth]{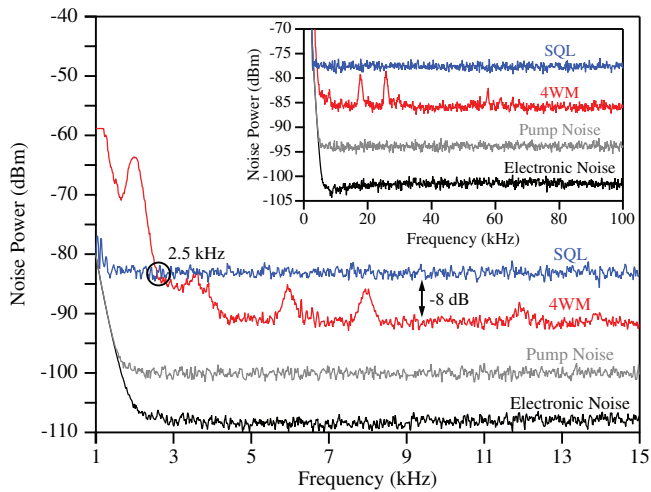}
    \caption{(color online). Low-frequency squeezing. Noise spectra
    (RBW=0.3~kHz, VBW=3~Hz) for the electronic noise, the pump
    scattering, the intensity-difference, and the standard-quantum
    limit. The inset shows a larger frequency span (RBW=1~kHz,
    VBW=10~Hz). \label{NoiseVFrequencyLow}} 
\end{figure}

In addition to looking at the level of noise reduction obtained at a
fixed frequency, we can also investigate the frequency spectrum of the
noise reduction.  Since there is no fundamental limitation on the
low-frequency response of the system~\cite{Lukin2000,Boyer2007}, it
will be established by technical noise on the pump and probe lasers.
The small number of optical components and particularly the lack of a
cavity minimizes the coupling to the environment. To explore this, we
record the intensity-difference noise spectrum at low analysis
frequencies with the detunings fixed at 800~MHz for the one-photon
detuning and 4~MHz for the two-photon detuning.  The probe (conjugate)
output powers are equal to 305 (290) $\mu$W, and the RBW and the VBW
are reduced (see Fig.~\ref{NoiseVFrequencyLow}). The
intensity-difference noise signal is 8.0~dB below the SQL and almost
flat, with the exception of a few resonance peaks, all the way down to
4.5~kHz.  At this point the technical noise of the pump and probe
lasers starts to dominate, resulting in the loss of the
intensity-difference squeezing at frequencies below 2.5~kHz. While
making these measurements we found that the frequency stabilization of
the Ti:sapphire laser adds amplitude noise to the beam, which in turn
prevents squeezing from being observed below 70~kHz. The data in
Fig.~\ref{NoiseVFrequencyLow} were taken with the active frequency
stabilization of the laser turned off. The observation of squeezing in
the kHz range makes our system suitable for applications such as the
transfer of optical squeezing onto matter waves~\cite{Haine2006}.

In addition to making the system insensitive to environmental noise,
the lack of a cavity allows the system to operate as a
multi-spatial-mode phase-insensitive amplifier~\cite{Boyer}, making
it an ideal source for quantum imaging
experiments~\cite{Kolobov2007}. When coupled to the low-frequency
squeezing capability of our system, multimode operation could find
an interesting application in photothermal spectroscopy, which
measures the deflection of a beam at frequencies of the order of
1~kHz, and is currently nearly limited by the
shot-noise~\cite{Owens1999}.

An important property of twin beams is the presence of
continuous-variable EPR (Einstein-Podolsky-Rosen)
entanglement~\cite{Reid1989}. In its current configuration, in which
the probe and conjugate are 6~GHz apart in frequency, the presence
of entanglement can be verified through the use of two different
local oscillators, or a bichromatic local oscillator
\cite{Marino2007}. In addition, the reciprocity between the beams
involved in the 4WM process should allow the pumping to occur at the
two frequencies $\omega_+$ and $\omega_-$, in order to generate
frequency degenerate twin beams at frequency $\omega$ or to realize a
phase-sensitive amplifier.

We have demonstrated a simple and robust source of
intensity-difference-squeezed light based on four-wave mixing in a
hot atomic vapor capable of producing a quantum noise reduction in
the intensity difference of more than 8~dB over a large frequency
range. The system provides a narrowband non-classical source near an
atomic transition and is well-suited for use in light-atom
interaction experiments. In addition, we have shown that under
certain conditions the system behaves as an ideal phase-insensitive
amplifier, opening the way to the generation of pure entangled
states. This realization of a high-quality source of non-classical
light may find a place in a variety of applications.

This work was supported in part by the IC postdoctoral program.  We
thank Luis Orozco and Ennio Arimondo for helpful discussions.

\end{document}